\documentclass[%
reprint,
superscriptaddress,
nobibnotes,
amsmath,amssymb,
aps,
prl,
]{revtex4-2}
\usepackage{soul}
\usepackage{braket,gensymb,tikz}
\usepackage{siunitx}
\usepackage{hyperref}
\usepackage{color}
\definecolor{orange}{RGB}{255,127,0}
\definecolor{blue2}{RGB}{33,114,173}
\usepackage{float}
\usepackage{xr}
\usepackage{graphicx}
\usepackage{xcolor} 

\usepackage{bm}

\begin{document}
	
\title{A conveyor-belt magneto-optical trap of CaF}

\author{Scarlett S. Yu}
\author{Jiaqi You}
\author{Yicheng Bao}
\altaffiliation[Current address: ]{Department of Physics, Princeton University, Princeton, NJ 08544, USA}
\affiliation{Department of Physics, Harvard University, Cambridge, MA 02138, USA}
\affiliation{Harvard-MIT Center for Ultracold Atoms, Cambridge, MA 02138, USA}

\author{Lo\"ic Anderegg}
\altaffiliation[Current address: ]{Department of Physics and Astronomy, University of Southern California, Los Angeles, CA 90089, USA}
\affiliation{Department of Physics, Harvard University, Cambridge, MA 02138, USA}
\affiliation{Harvard-MIT Center for Ultracold Atoms, Cambridge, MA 02138, USA}

\author{Christian Hallas}
\author{Grace K. Li}
\affiliation{Department of Physics, Harvard University, Cambridge, MA 02138, USA}
\affiliation{Harvard-MIT Center for Ultracold Atoms, Cambridge, MA 02138, USA}

\author{Dongkyu Lim}
\author{Eunmi Chae}
\affiliation{Department of Physics, Korea University, Seongbuk-gu, Seoul 02841, South Korea}

\author{Wolfgang Ketterle}
\affiliation{Harvard-MIT Center for Ultracold Atoms, Cambridge, MA 02138, USA}
\affiliation{Department of Physics, Massachusetts Institute of Technology, Cambridge, MA 02139, USA}

\author{Kang-Kuen Ni} 
\affiliation{Department of Physics, Harvard University, Cambridge, MA 02138, USA}
\affiliation{Harvard-MIT Center for Ultracold Atoms, Cambridge, MA 02138, USA}
\affiliation{Department of Chemistry and Chemical Biology, Harvard University, Cambridge, MA 02138, USA}

\author{John M. Doyle} 
\affiliation{Department of Physics, Harvard University, Cambridge, MA 02138, USA}
\affiliation{Harvard-MIT Center for Ultracold Atoms, Cambridge, MA 02138, USA}

\date{\today}

\begin{abstract}
 We report the experimental realization of a conveyor-belt magneto-optical trap for calcium monofluoride (CaF) molecules. The obtained highly-compressed cloud has a mean radius of 64(5)~$\mu$m and a peak number density of $3.6(5) \times 10^{10}$~cm$^{-3}$, a 600-fold increase over the conventional red-detuned MOTs of CaF, and the densest molecular MOT observed to date. Subsequent loading of these molecules into an optical dipole trap yields up to $2.6 \times 10^4$ trapped molecules at a temperature of 14(2)~$\mu$K with a peak phase-space density of $\sim 2.4 \times 10^{-6}$. This opens new possibilities for a range of applications utilizing high-density, optically trapped ultracold molecules.
	\end{abstract}

\maketitle

 The diverse quantum phenomena and potential applications uniquely offered by ultracold molecules are owed to their intrinsic electric dipole moment and intricate internal structures. The promise of new science with molecules has driven significant advances in direct laser cooling and magneto-optical trapping of molecules in the past decade. Applications range from quantum simulation \cite{micheli2006toolbox, gadway2016strongly, baranov2008theoretical,wall2015magnetism, gorshkov2011tunable, cornish2024quantum} and computation \cite{demille2002quantum, yelin2006dipolarQC, karra2016paramagnetic, sawant2020qudits, ni2018dipolar}, to precision measurement tests for fundamental physics \cite{doyle2022ultracold, kozyryev2017precision, kozyryev2021enhanced, hutzler2020polyatomic, anderegg2023quantum, norrgard2019nuclear, hao2020nuclear}, and studies of quantum-state-controlled chemistry and collisions \cite{heazlewood2021towards,bell2009ultracold, anderegg2021observation}.
 The feasibility and further improvement of many of the current and proposed studies of cold molecules hinge on having a high number of trapped ultracold molecules.
For example, in shot-noise limited precision measurements, such as those probing the electron electric dipole moment, the statistical uncertainty decreases with increasing number of trapped molecules $N$ as $\sim 1/\sqrt{N}$. With smaller and denser molecular clouds, loading molecules into an optical dipole trap becomes more efficient, thereby increasing $N$ and sensitivity for experiments \cite{hutzler2020polyatomic}.
Similarly, to perform evaporative cooling towards the creation of quantum degenerate gases of ultracold molecules, a high initial density is necessary to facilitate rapid elastic collisions to occur, as well as to retain sufficient number of molecules during the cooling process \cite{anderson1995observation, schindewolf2022evaporation, son2020collisional}. Finally, high densities are also desired in ultracold molecular collisional studies \cite{son2020collisional}.

The magneto-optical trap (MOT) is a pivotal technique for generating cold, dense samples of atomic gases, and has now been progressively extended to diatomic \cite{collopy20183d, anderegg2017radio, norrgard2016submillikelvin, zeng2024three} and polyatomic molecular gases \cite{vilas2022magneto,lasner2024magneto}. It serves as an essential starting point for further cooling and subsequent transfer into conservative optical potentials~\cite{anderegg2019optical, anderegg2018laser, hallas2023optical, jorapur2024high, bao2024Raman}. Molecular MOTs are conventionally formed using red-detuned light (``red MOT"), where the trapping light is tuned below the transition resonance. These red MOTs typically contain $10^4-10^5$ molecules, with a peak number density of $ \sim 10^5-10^6 \text{ cm}^{-3}$ and phase-space density of $\sim 10^{-12}-10^{-14}$ \cite{anderegg2017radio, collopy20183d, langin2023toward}.
These values are roughly six orders of magnitude lower than those routinely reached in atomic MOTs. 
Laser cooling of alkali atoms is typically realized on type-I transitions with $F'> F$, where $F(F')$ is the quantum number of total angular momentum of ground (excited) electronic state. In contrast, all laser cooling of molecules to date operates on type-II transitions ($F' \leq F$) to ensure rotational closure for effective optical cycling \cite{devlin2016three, devlin2018laser, collopy20183d, tarbutt2015magneto}. 
Studies \cite{devlin2016three} have shown that type-II red MOTs experience sub-Doppler heating at low velocities, producing a higher temperature and lower density cloud that has a limited overlap volume with the tightly-confined optical dipole trap (ODT).  
This, combined with a low molecule number in the MOT due to insufficient laser slowing of molecular beams in general, 
results in only a few thousands of trapped molecules in the optical dipole trap (ODT) and a phase-space density prohibitively low to realize many potential applications \cite{anderegg2019optical, langin2021polarization, hallas2023optical}. 
However, numerical simulations \cite{jarvis2018blue} have revealed that for a type-II system in a 3D molasses, the sub-Doppler Sisyphus-type force near zero velocity persists over the range of magnetic fields typically seen in the MOT. This finding motivated the introduction of a blue-detuned MOT (``blue MOT") stage, utilizing sub-Doppler processes in the presence of a magnetic field and implemented after the red MOT to increase compression and, consequently, phase-space density. This approach, first demonstrated with Rb atoms \cite{jarvis2018blue}, has since been successfully implemented on several laser-cooled diatomic molecules, consistently producing substantially colder and denser samples \cite{jorapur2024high, burau2023blue, li2024blue} with peak number densities of $ n_o \approx 10^7 - 10^8 \text{ cm}^{-3}$ and phase-space density of $\rho \approx 10^{-9}$. The reported cloud mean radii were $\gtrsim \si{150}{\mu m}$, limited by position-dependent trapping forces. These limitations prompted the recent development of a blue MOT \cite{hallas2024high, Grace_bMOT} using a frequency scheme that produces an additional physical effect, generating significantly more effective compression than the previously demonstrated blue MOTs. The improvement in the trapping force is attributed to Sisyphus-type blue-detuned cooling in the presence of a magnetic field gradient occurring in combination with a slowly-moving optical molasses  \cite{hallas2024high, Grace_bMOT}. This ``conveyor-belt" effect shuttles the molecules towards the center region of the MOT and deposits them there.

\begin{figure}[!ht]
		\centering
            \includegraphics[width=0.9\linewidth]{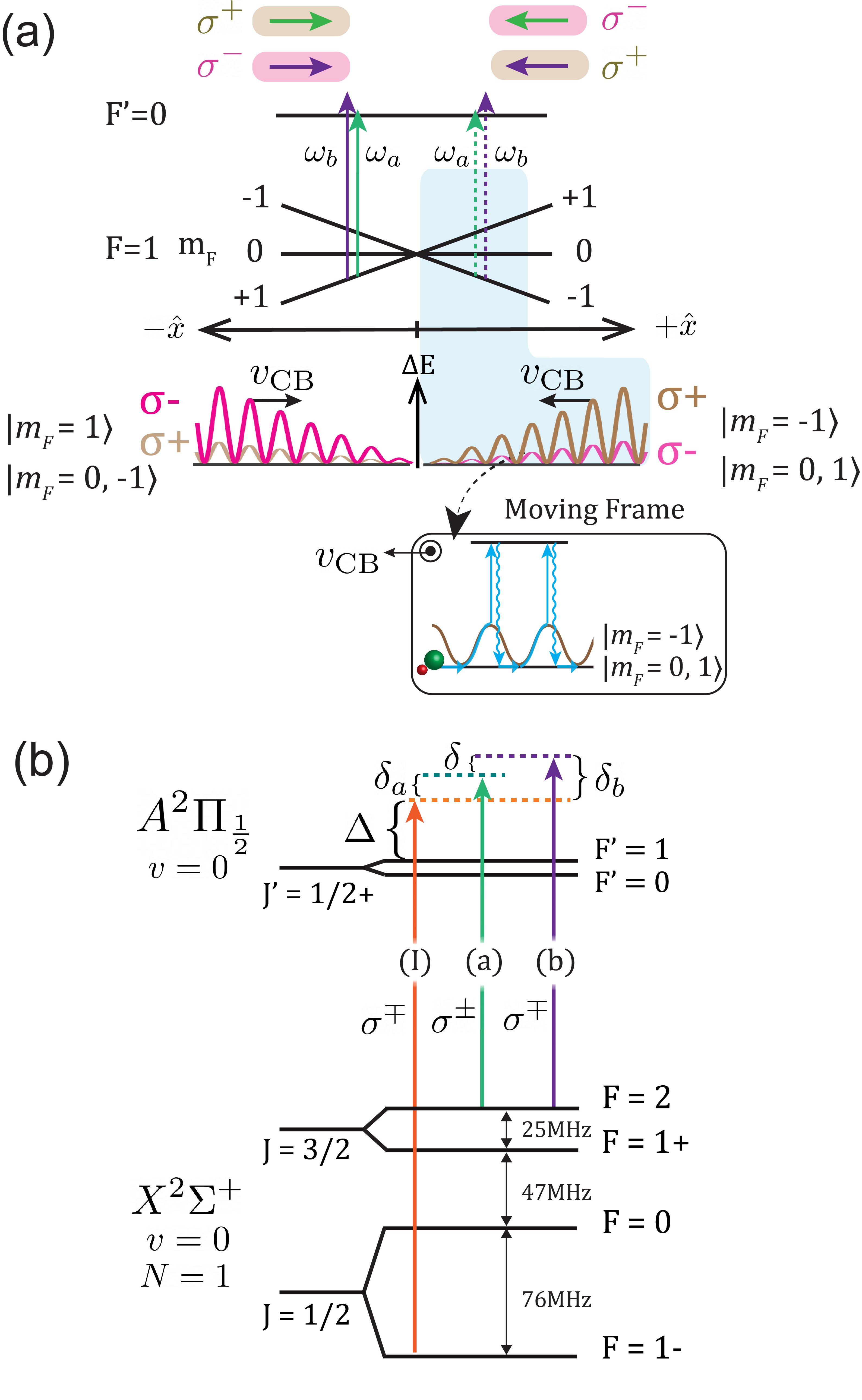}
		\caption{(a) Illustration of the working principle of a conveyor-belt blue-detuned MOT; see text. The energy scale is shown for illustration purposes only. (b) Relevant level structure of CaF and laser configurations used in this work. J and F label the hyperfine states, where $J$ the total angular momentum without nuclear spin and $F$ is the total angular momentum. }
		\label{fig:1}
\end{figure}

In this letter, we experimentally realize a conveyor-belt blue-detuned MOT (simply, a conveyor MOT) for calcium monofluoride (CaF) molecules, capturing $N_\text{MOT}\simeq 1.3\times10^{5}$ molecules in a mean spatial radius of $\SI{64(5)}{\mu m}$ and reaching a peak number density of $n_o^\text{MOT} =3.6(5) \times \si{10^{10}}{\text{ cm}^{-3}}$, which is a more than 600-fold increase over the compressed red-detuned MOT, and the densest molecular MOT observed to date. We subsequently load the molecules into an optical dipole trap, trapping up to $N_\text{ODT} \simeq ~2.6 ~\times ~10^4$ molecules at temperatures of $T_\text{ODT}$ = ~14(2)~$\mu$K, achieving a peak number density $n_o^\text{ODT} = 1.05 \times 10^{10}$~cm$^{-3}$ and a phase-space density as high as $\rho_{\text{ODT}} = 2.4 \times 10^{-6}$. This $N_\text{ODT}$ represents a 170-fold increase over the previously reported number for CaF molecules~\cite{anderegg2019optical} and, to the best of our knowledge, is the highest number of directly laser-cooled molecules in an conservative optical potential reported thus far. Our results align with and provide strong support for the enhanced trapping mechanism of the conveyor MOT, initially proposed and observed in~\cite{hallas2024high} for CaOH molecules and theoretically developed in~\cite{Grace_bMOT}. The molecule number density achieved here is comparable to those typically obtained in atomic optical dipole traps, providing a promising starting point of numerous near-term applications that reply on high densities of optically trapped ultracold molecules.

The underlying working principle of the conveyor MOT, illustrated in Fig.~\ref{fig:1}(a) and supported by numerical simulations~\cite{Grace_bMOT}, is as follows. Consider a MOT beam containing two closely-spaced frequency components, $\omega_{a}$ and $\omega_{b}$, each with opposite circular polarizations ($\sigma^{+}$ and $\sigma^{-}$, respectively). As in the conventional MOT configuration, the beams are directed towards the MOT center along, say, the $\hat{x}$ direction, and retro-reflected, with the retro-reflected beam containing the same frequency components but flipped polarizations. When two counter-propagating beams of the same polarization interfere, they form a standing wave. If these beams have a relative detuning $\delta$, the resultant wave will move along $\hat{x}$ in the lab frame, creating the conveyor belt effect, much like a moving lattice.  Here, the $\omega_{a}$ ($\sigma^{+}$) light travelling towards the $+\hat{x}$ and the $\omega_{b}$ ($\sigma^{+}$) towards $-\hat{x}$ will form a ``$\sigma^{+}$ conveyor belt" that slowly travels along $-\hat{x}$ direction at a velocity $v_{\text{CB}} \approx \frac{\delta}{2  \omega_{a}}c$ where $ \delta = \omega_{b} - \omega_{a}$. Similarly, a ``$\sigma^{-}$ conveyor belt" moving in $+\hat{x}$ is formed by $\omega_{a}$ ($\sigma^{-}$) travelling towards $-\hat{x}$ and $\omega_{b}$ ($\sigma^{+}$) in $+\hat{x}$.

Now consider a type-II transition between hyperfine levels $F=1 \leftrightarrow F'=0$, where the ground-state has a nonzero $g$-factor $g_F$ (assume, without loss of generality, that $g_F >0$). The presence of magnetic field gradients creates a position-dependent interaction strength between the molecule with either the $\sigma^{+}$ or $\sigma^{-}$ conveyor belts, depending on the sign of $g_F$. If $\omega_{a}$ and $\omega_{b}$ in the MOT beam are blue-detuned from the transition, in the $B > 0$ region the $|F=1, m_F =-1\rangle$ magnetic substate is Zeeman-shifted closer to resonance and thus molecules preferentially interact with the $\sigma^{+}$ conveyor belt. In the conveyor belt's moving frame, the $\sigma^{+}$ light appears to the molecules as a stationary wave, setting up a spatially periodic light shift to facilitate Sisyphus-type cooling, where molecules experience maximum optical pumping rate as it move across the intensity maxima, before spontaneously decaying to the $|F=1, m_F =0,1\rangle$ dark states and completing a cooling cycle~\cite{devlin2016three}. In the lab frame, the molecules get accelerated towards $\sim v_{\text{CB}}$ and transported towards the MOT center by the moving $\sigma^{+}$ conveyor belt. On the other side of the MOT, in the $B < 0$ region, the molecules preferentially interact with the $\sigma^{-}$ conveyor belt while experiencing Sisyphus cooling in the $\sigma^{-}$ conveyor belt's moving frame. Near the MOT center where the magnetic field is weak, the preferential interaction with the $\sigma^{\pm}$ conveyor belts diminishes. At the center, the dominant cooling is provided by gray-molasses cooling in the lab frame with $\Lambda$-systems forming between the hyperfine levels. This, in part, effectively prevents the molecules from being over-transported out of the MOT center region. In summary, in a conveyor MOT, molecules away from the MOT center on either sides selectively interact with one of the $\sigma^{\pm}$ conveyor belts, in which they are cooled into the moving frame of the conveyor belt and slowed to near zero velocity as they approach the center, resulting in a large number of near zero-velocity molecules piled up in the MOT center.

In our experiment, CaF molecules are initially captured in a radio-frequency (RF) red MOT operating on the $X(\nu=0, N=1) - A(\nu'=0, J'=1/2)$ cooling transition \cite{anderegg2017radio}. The apparatus and optical cycling scheme are described in our previous work \cite{anderegg2017radio,bao2022fast,bao2023dipolar}. 
We begin with a magnetic field gradient of $B^{'}=$~20~G/cm RMS and a total intensity of $I_o= 30$~mW/cm$^2$ per MOT beam at a detuning of $\Delta = 2 \pi \times -8$~MHz and load the MOT for 5~ms. The intensity is then reduced to $I_o/4$ in 5~ms and the magnetic field gradient is ramped to $B' =$ 50~G/cm RMS, compressing the molecular cloud to $\sigma \sim$~600~$\mu$m, with $2 \times 10^5$ trapped molecules, corresponding to $n_o^{\text{red}}\approx ~5.8 \times 10^7\text{~cm}^{-3}$. In the next 5 ms, we turn off the RF magnetic field and polarization-switching, and jump the laser intensity and detuning to the free-space $\Lambda$-cooling optical scheme \cite{anderegg2018laser, cheuk2018lambda}. A subsequent 5-ms sub-Doppler cooling pulse reduces the temperature of the molecular cloud to 18(5)~$\mu$K while retaining $\sim 80\%$ of the initially captured molecules. 

Next, we load the molecules from the red MOT into a DC conveyor MOT. To do so, we first switch the MOT light frequency to the configuration shown in Fig.~\ref{fig:1}(b) by introducing one additional frequency component to form the conveyor MOT. In this configuration, the frequency $\omega_{I}$, addressing the $|F=1-\rangle$ hyperfine state, is blue-detuned with a single-photon detuning $\Delta$. The other two frequency components, $\omega_{a}$ and $\omega_{b}$, both addressing the $|F=2\rangle$ state, have opposite circular polarizations and detunings of $\Delta+\delta_{a}$ and $\Delta+\delta_{b}$, respectively. These two closely-spaced components with a slight relative detuning of $\delta = \delta_b - \delta_a$ create the conveyor-belt effect and consequently an enhanced trapping force on the $|F=2\rangle$ state, while $\omega_{I}$ mainly functions as a repumper and facilitates lab-frame cooling at the MOT center. We then gradually increase the DC magnetic field using an optimized profile that minimizes molecular loss. 
We realize the switch from RF to DC MOT by employing high-impedance inductors and back-to-back solid-state relays. See Supplemental Materials \cite{SM}. Finally, to detect and measure the spatial profile of the molecular cloud, we collect in-situ fluorescence onto a camera.

\begin{figure}[!ht]
		\centering
		\includegraphics[width=1\linewidth]{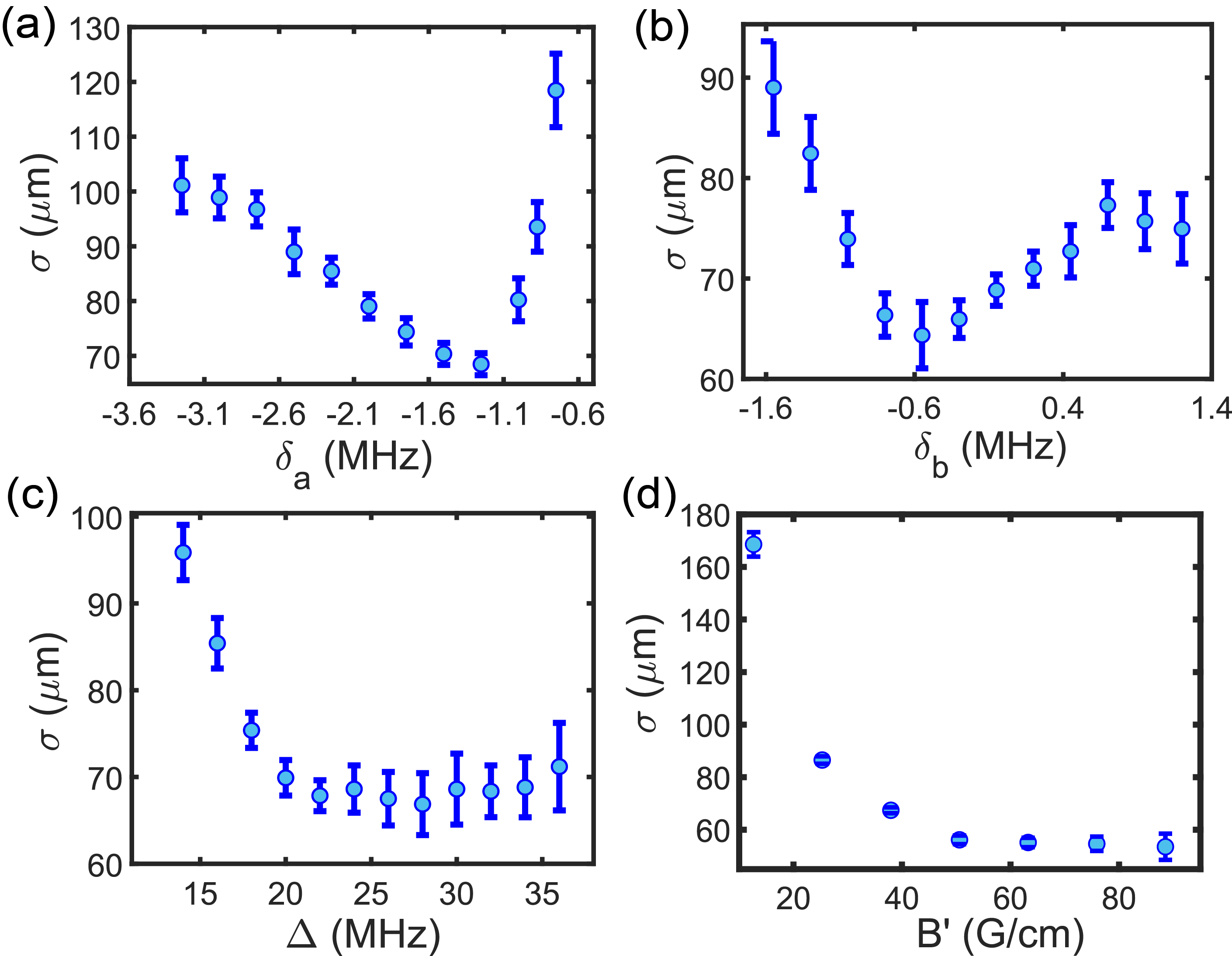}
		\caption{Dependence of conveyor MOT cloud size on frequency detunings and magnetic field. (a) MOT radius $\sigma$ as a function of two-photon detunings (a) $\delta_{a}$, (b) $\delta_{b}$, (c) single-photon detuning $\Delta$, and (d) applied magnetic field gradient $B^{'}$. For each scan in (a-c), all other detunings and the total intensity are kept fixed. The intensity of each beam is calibrated such that it remains constant as frequency is varied. The maximum magnetic field gradient used in (a-c) is $B^{'}\approx40$~G/cm.} 
		\label{fig:2}
\end{figure}

To maximize compression of the molecular cloud, we investigate the effect of single and two-photon detunings ($\Delta$, $\delta_{a}$ and $\delta_{b}$) on the mean radius $\sigma$ of molecules in the conveyor MOT. Fig. \ref{fig:2}(a-c) show $\sigma$ as a function of $\Delta$, $\delta_{a}$ and $\delta_{b}$. We obtain $\sigma$ in each direction by fitting it to a Gaussian distribution. $\sigma$ is minimized at $\delta_{a}= - 2\pi \times 1.5$~MHz, $\delta_{b}= 2\pi \times 0.5$~MHz, corresponding to a conveyor belt speed of approximately $v_{\text{CB}}\approx 0.6$~m/s. 
Physically, due to the decreasing magnetic field as the molecules are compressed towards MOT center, combined with the frictional force imparted by the opposing conveyor belt, the molecules on average do not get fully accelerated to $v_{\text{CB}}$. $\sigma$ is relatively insensitive to $\Delta$ over the range $\Delta = 2\pi \times$ (20~MHz to 30~MHz), similar to the observations in \cite{hallas2024high}. This detuning range suggests that part of the trapping force may also originate from the $|F=1+\rangle$ state. $\Delta =20$~MHz is used for all subsequent measurements.
In Fig.~\ref{fig:2}(d), a scan of $B^{'}$ reveals that $\sigma$ =~64(5)~$\mu$m (with radial and axial widths of $\sigma_{\text{ax}}$~=~58(4)~$\mu$m and $\sigma_{\text{rad}}$ =~68(2)~$\mu$m, respectively) is reached at $B^{'}\approx40 ~\text{G/cm}$. The light intensity used for the conveyor MOT 
is $\sim 2 I_\text{sat}$ and the total light power is equally distributed among the three frequency components. We observe that the minimum $\sigma$ is relatively insensitive to the ratio of power intensities for the three components. With the available laser power in our experiment, we observe an increase in the number of trapped molecules with higher power (Fig. S4), as expected from the corresponding enhancement in both capture velocity and trap volume for gray-molasses cooling.
\begin{figure}[!ht]
		\centering
		\includegraphics[width=1\linewidth]{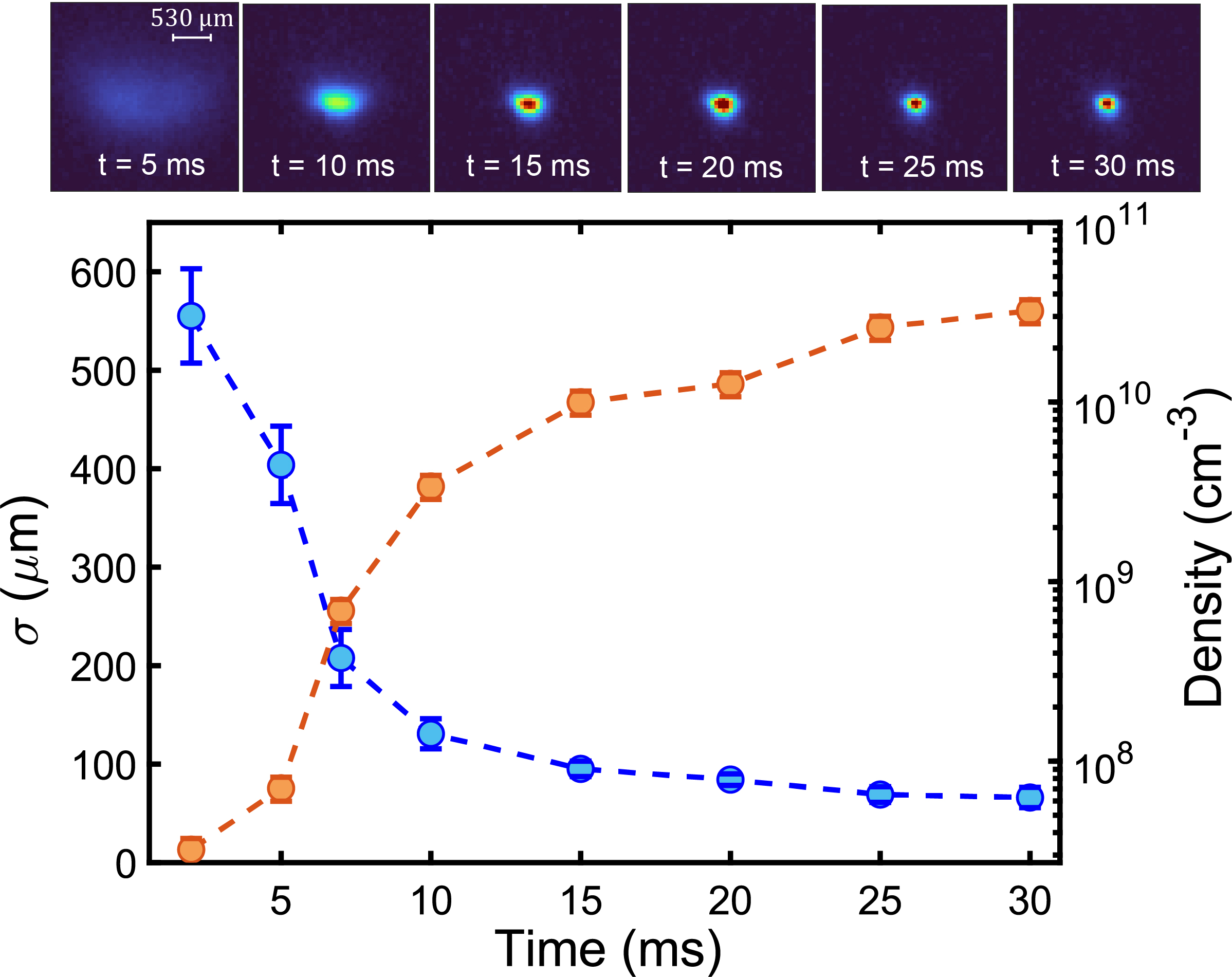}
		\caption{Fitted cloud radius $\sigma$ (left axis) and peak molecule number density $n_o$ (right axis) of the conveyor-belt assisted blue MOT as a function of time (dashed lines are guides to the eye). Data are taken with frequency parameters $\Delta = 2\pi\times 20$~MHz, $\delta_a =-2\pi\times 1.5$~MHz and $\delta_b = 2\pi\times 0.5$~MHz, and a magnetic field gradient of up to $B^{'}\approx40$~G/cm. Top: Images of the molecular cloud taken by collecting 2 ms of in-situ fluorescence following a compression time of $t$.}
		\label{fig:3}
\end{figure}

Fig.~\ref{fig:3} presents $\sigma$ and $n_o$ as a function of conveyor MOT application time, with top images showing the in-situ fluorescence. Most of the compression takes place within the first $10$~ms, during which the cloud reduces from $\sigma \gtrsim 550 \mu$m to $\sigma= 130~\mu$m as $B^{'}$ is ramped to $\sim$~20~G/cm. Over the next 20~ms, $\sigma$ continues to decrease as $B^{'}$ reaches 40~G/cm.
By the end of the compression, up to $N_\text{MOT}\sim 1.3\times10^5$ molecules remain trapped, corresponding to a peak density of $n_o^\text{MOT} =~3.6(5)\times10^{10}$~cm$^{-3}$, with measured mean temperature of $T_\text{MOT}=$ 141(7)~$\mu$K.  
Given the relatively low measured photon scattering rate of $0.27(3)\times 10^6~\text{ s}^{-1}$ in the conveyor MOT (using laser parameters optimized for minimal $\sigma$) (Fig. S2), it may be that higher light intensity could provide larger trapping volume, but this is open for further investigation.  
In addition, we find that the $1/e$ lifetime of molecules in the conveyor MOT follows a single exponential decay, with a fitted $\tau= 37.2(8)$~ms measured for $B^{'} \approx 40$~G/cm (Fig. S1), consistent with the absence of light-assisted two-body collisions at this density, which can often limit the density in atomic MOTs. 

\begin{figure}[!ht]
		\centering
		\includegraphics[width=1\linewidth]{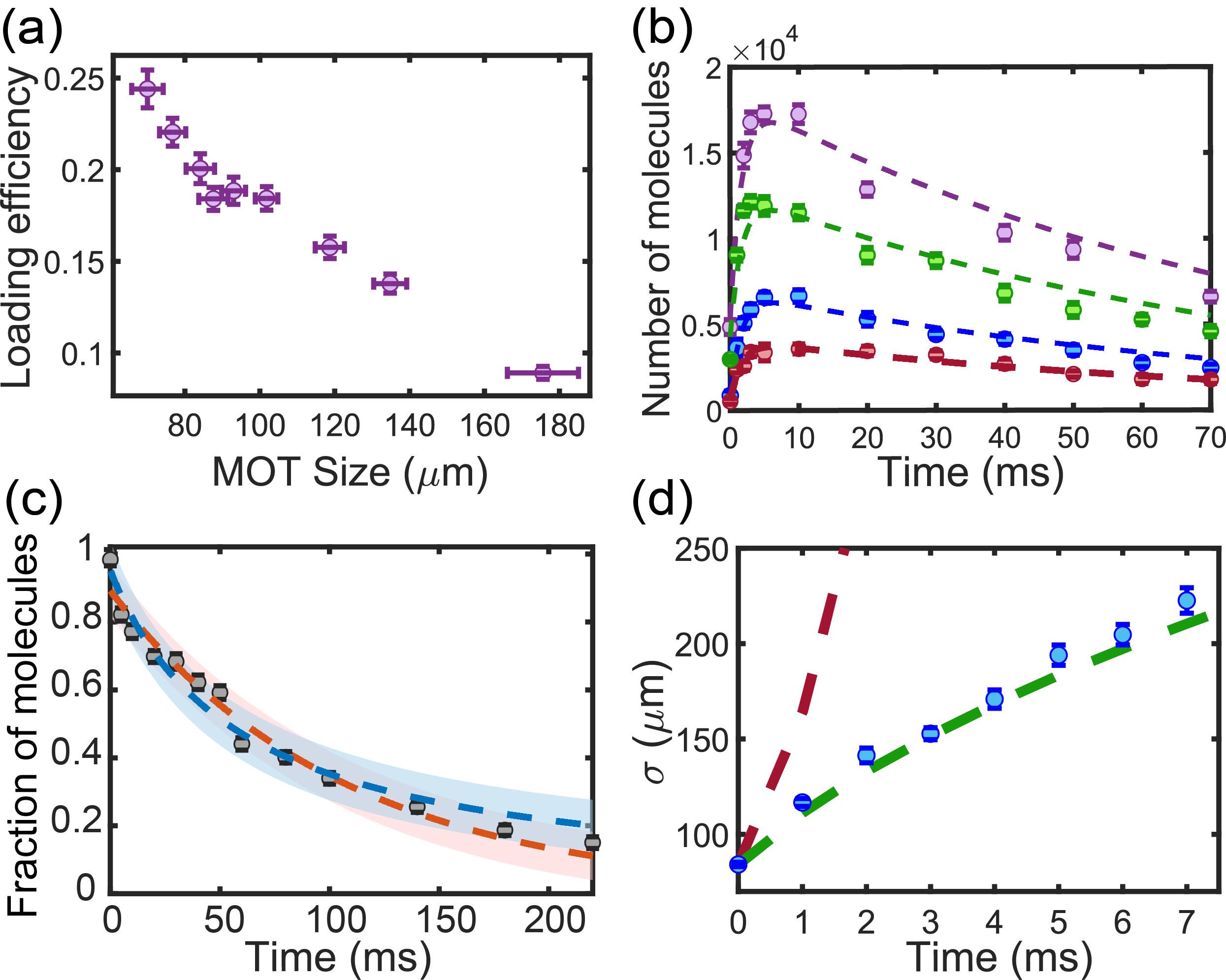}
		\caption{ODT loading. (a) Loading efficiency as a function of conveyor MOT cloud size. (b) Number of molecules in the ODT as a function of loading time for different initial number of molecules. (c) Fraction of molecules remaining in the ODT as a function of hold time in presence of $\Lambda$-cooling light. (d) Expansion of the reservoir cloud in presence of $\Lambda$-cooling as a function of time. The ballistic expansion trajectory for cloud at 147~$\mu$K is plotted as a guide (red, dashed). A diffusive model fitted to the data is shown (green, dashed). }
		\label{fig:4}
\end{figure}

The high molecular density obtained in the conveyor MOT provides a reservoir from which we transfer molecules into a 1064~nm ODT of 57~$\mu$m beam waist and a trap depth of $k_B \times U \approx$ 174~$\mu$K. To load the ODT, we switch off all MOT light and DC magnetic field, and immediately turn on the ODT light in the presence of $\Lambda$-cooling light for 3 ms. 
The number of molecules transferred into the ODT depends on $\sigma$ of the cloud reservoir as expected, as shown in Fig.~\ref{fig:4}(a). The loading efficiency increases as the size of the conveyor MOT cloud approaches that of the ODT, leading to better spatial mode-matching between the two. Up to $24 \%$ of the molecules from the cloud reservoir are captured into the optical trap, resulting in $N_\text{ODT} \simeq 2.6 \times 10^4$ molecules at $T_\text{ODT}=$ 14(2)$~\mu$K with $\rho_{\text{ODT}} = 2.4\times10^{-6}$. This corresponds to $12 \%$ of the molecules initially trapped in the red MOT ending up in the ODT, a factor of five improvement over our previous loading efficiency of $\lesssim 2\%$ using direct transfer from the red MOT. 

To understand the possible loss mechanisms during the loading process, we examine the time dependence of the molecular number during the ODT loading. Fig.~\ref{fig:4}(b) shows the loading curves of $N_\text{ODT}$ at varying $N_\text{MOT}$ in the cloud reservoir, $N_\text{MOT}$. The ODT loading dynamics are well described by a rate equation model~\cite{PhysRevA.62.013406} by accounting for overall molecular loss in the reservoir cloud, one-body, and two-body loss mechanisms (see Supplemental Materials~\cite{SM}). To determine whether two-body collisional loss plays a role in limiting the loading, we measure the lifetime of molecules in the ODT in presence of varying $\Lambda$-light duration \cite{cheuk2020observation, anderegg2021observation}.  
Fitting the data to two-body loss model yields a decay rate that is statistically indistinguishable from that obtained using a single-body loss model (Fig.~\ref{fig:4}(c)), indicating negligible two-body loss in this system. Additionally, the linear scaling observed between $N_\text{ODT}$ and $N_\text{MOT}$ suggests that current loading is far from saturation. 

The results in Fig.~\ref{fig:4}(d) reveal that the effective ODT loading time, as seen in Fig.~\ref{fig:4}(b), is limited by the rapid expansion of the molecular cloud. $\sigma$ increases to 153~$\mu$m in the first 3~ms in the presence of $\Lambda$-cooling light, leading to a large spatial mismatch with the ODT. With a typical $\Lambda$-cooling time of $< 300~\mu$s, the expansion quickly deviates from a ballistic trajectory that would be expected for the 147~$\mu$K conveyor MOT temperature, instead following a diffusion trajectory (dashed lines in Fig.~\ref{fig:4}(d)). 
This rapid diffusive expansion is the primary limitation on our loading efficiency, and larger optical dipole trap volume would likely improve it. In the future, the cloud expansion issue might be addressed by directly loading the ODT from the conveyor MOT, provided the optical trap depth is much higher than the conveyor MOT cloud temperature.  

In summary, we realize a high density  MOT of CaF using the conveyor-belt mechanism. 
Our work experimentally strongly validates the general nature and the effect of the conveyor-belt trapping mechanism initially seen in \cite{hallas2024high}. The conveyor MOT provides strong trapping forces for creating highly compressed molecular samples, which subsequently enables us to efficiently load the molecules into an optical dipole trap within the expansion time of the reservoir cloud. As a result, approximately $2.6 \times 10^4$ molecules are loaded into an optical dipole trap at a phase-space density of $2.4\times 10^{-6}$. Having a high molecular number and phase-space density in optical traps expands the possibilities for applications in precision measurement using ultracold molecules and evaporative cooling for creating a quantum degenerate gas of laser-cooled molecules. This effective and straightforward conveyor MOT should be readily applicable to other molecular MOTs, as well as type-II atomic systems \cite{Grace_bMOT}.

\section*{Acknowledgment}
This material is based upon work supported by the U.S. Department of Energy, Office of Science, National Quantum Information Science Research Centers, Quantum Systems Accelerator and Harvard-MIT Center for Ultracold Atoms (Grant No. PHY-2317134). Additional support is provided by AFOSR, AOARD, and ARO. S.Y. acknowledges support from the NSF GRFP. L.A. and S. Y. acknowledge support from the HQI. E.C. acknowledges support from the NRF of Korea (Grants No. 2022M3C1C8097622, RS-2024-00439981, and RS-2024-00531938).

\bibliography{caf_bmot.bib}

\end{document}